\documentstyle[twocolumn,aps,epsfig,floats]{revtex}
\begin{document}
\draft
\title{Transverse electrokinetic and microfluidic
effects in micro-patterned
channels:\\
lubrication analysis for slab geometries}
\author{Armand Ajdari}
\address{Laboratoire de Physico-Chimie Th\'eorique, Esa CNRS 7083, ESPCI,
  10 rue Vauquelin, F-75231 Paris Cedex 05, France\\
\rm {submitted to Phys. Rev. E. dec. 2000}}
\address{
\begin{minipage}{5.55in}
\begin{abstract}\hskip 0.15in 
Off-diagonal (transverse) effects in micro-patterned geometries
are predicted and analyzed within the general frame of linear response theory,
relating applied
presure gradient and electric field to flow and electric current.
These effects could contribute to the design of
pumps, mixers or flow detectors.
Shape and charge density modulations are proposed as a means
to obtain sizeable transverse effects, as demonstrated by
focusing on simple geometries and using
the lubrication approximation.
\end{abstract}
\end{minipage}
\vspace*{-0.5cm} 
}
\maketitle




\newpage
\section{Introduction}

The development of microfluidic devices and studies
has prompted the quest for various strategies to achieve
pumping in micro-geometries \cite{Mtas1,Str2,Fuh1,Mul1,Was1,Gal1,Har1}.
Pressure driven flows is the first obvious possibility,
with the inconvenience of important Taylor dispersion
due to the parabolic flow profiles \cite{Rus}.
Electro-osmosis has been proposed and developed as a way to generate
almost perfect plug-flows, thereby reducing dispersion
in various devices, 
which results in limited dilution of samples, and processability
for separation purposes \cite{Har1}. This solution
implies relatively high voltages
applied between the ends of the channels.

In this paper, micro-patterned
channels are proposed as a way to generate a large class of effects
using pressure gradients
or electric fields.
In particular various off-diagonal
effects can be obtained in which a cause along one direction
leads to a measurable or useful effect in a perpendicular direction.
These effects could be exploited for the realization
of transverse pumps, mixers, flow detectors, etc ...
A proposed pattern is the periodic modulation
of the shape of the channels,
which can be improved by a combined modulation
of the surface charge density.
(in line with an earlier study that dealt with transverse electro-osmosis on
such surfaces \cite{Ajd1}).
The aim is here
to explore the ensemble of transverse effects achievable.

To reach this goal, the linear response regime is considered,
which permits the use of a very general Onsager formalism, 
to relate fields (pressure and electric potential gradients) to 
fluxes (hydrodynamic flow and electric current).
For the sake of clarity, we further restrain our
analysis to a simple class of geometries,
with the fluid confined between two parallel plates, 
homogeneously and periodically patterned (Fig. 1).
Parallel walls may be present so as to
form a capillary.

The geometry and
the matrix representation of the linear Onsager formalism are proposed in Section II.
In section III the off-diagonal
effects are described at a phenomenological level,
for the case of passive walls (a few situations where electrodes are embedded
in the walls are considered in Appendix 1, and the corresponding
wall generated effects analyzed).
In section IV an explicit
realization is described and 
estimates for the effects given. The calculations
are performed using the lubrication approximation,
and assuming weak surface potentials,
which gives a convenient analytic handle on
these complex systems. Results
for sinusoidal modulations of the shape and charge densities
are reported in Appendix 2.
A brief discussion closes the paper (section V), pointing out directions for future studies.
\begin{figure}
\vspace{-.5cm}
\centerline{\epsfig{figure=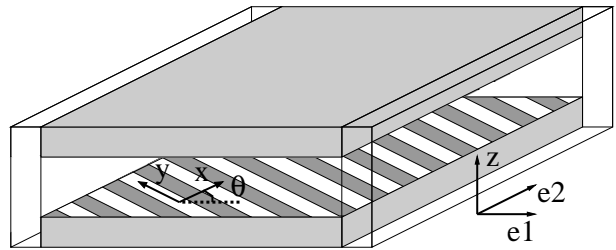}}
\vspace{.5cm}
\caption{
Slab geometry considered in this paper: Two parallel walls bear on their
inner faces periodic micro-fabricated patterns of principal axes $x$ and $y$.
The slab can be limited sideways by walls, the axis of the resulting capillary
$e_2$ at an angle $\theta$ with the pattern.
From section IV on, we focus on patterns periodic along
$x$ and invariant along $y$. 
}\label{f_1}\end{figure}
\section{Local linear response}
We consider the Hele-Shaw quasi-planar geometry of Figure 1,
where the two sides of the slab bear patterns 
of principal axes $(x,y)$, and set the formalism relating
locally 2D fields to 2D fluxes (i.e. integrated over $z$).
\subsection{In the principal axes $(x,y)$}
In the linear response regime,
the 2D currents (hydrodynamic flow ${\bf J}$
and electric current ${\bf J_{el}}$)
are related to the pressure gradient ${\bf \nabla}p$
and to the electrostatic potential gradient ${\bf \nabla}\phi=-{\bf E}$,
by a matrix equation \cite{Hun1}:
\begin{eqnarray}
{\bf J} &=& {\bf \hat K.}(-{\bf \nabla}p) + {\bf \hat M.}(-{\bf \nabla}\phi)
                \label{Toh} \\
{\bf J_{el}} &=& {\bf \hat M.}(-{\bf \nabla}p) + {\bf \hat
S.}(-{\bf \nabla}\phi).
                \label{Tih}
\end{eqnarray}
where the matrices ${\bf \hat F}$ with $F$ equal to $K$, $M$ or $S$
are diagonal in the $(x,y)$ basis: 
\begin{equation}
{\bf \hat F}=\left( \begin{array}{ll}
F_x & 0 \\
0 & F_y  \end{array} \right)  
\end{equation}
The permeation matrix ${\bf \hat K}$ describes the flow 
induced by pressure differences (the Darcy law in
a porous medium). The conduction matrix ${\bf \hat S}$ 
relates the electric current to the electric field (the medium's Ohm's law).
The matrix ${\bf \hat M}$ describes the electro-hydrodynamic coupling
(so-called electrokinetic effects).
In equation (\ref{Toh}) it quantifies {\em electro-osmosis}, i.e. the
hydrodynamic flow induced by the electric field.
This effect stems from the presence of
thin diffuse layers in the vicinity
of charged walls where the fluid is non-neutral, and thus dragged
by the local electric field. 
In (\ref{Tih}) ${\bf \hat M}$ measures the electric current induced 
by the presence of a net hydrodynamic flow. This is due to
the convective transport of the above mentioned charged layer
which leads to an ionic current. The result is hydrodynamically generated
electrostatic potential differences (``streaming potentials'')
and electric currents (``streaming currents'').

Equivalently, the inverse system (the resistance matrix),
gives the fields as functions of the currents:
\begin{eqnarray}
-{\bf \nabla}p &=& {\bf {\hat k}.J} + {\bf {\hat m}.J_{el}}
                \label{to} \\
-{\bf \nabla}\phi &=& {\bf {\hat m}.J} + {\bf 
{\hat s}.J_{el}} .
                \label{ti}
\end{eqnarray}
where the matrices ${\bf {\hat k}}$, ${\bf {\hat m}}$ and ${\bf {\hat s}}$,
are diagonal 2x2 matrices of the form (3) with, 
 for $i= x,y$ :
\begin{eqnarray}
k_i &= S_i/\Delta_i\, ,\, m_i &= -M_i /\Delta_i \\ 
s_i &= K_i/ \Delta_i \, ,\,
\Delta_i &= K_iS_i-M_i^2
\end{eqnarray}
\subsection{In an arbitrary $(e_1,e_2)$ coordinate system}
All the above is standard in the 1D geometry 
of cylindrical capillaries, or for homogeneous and isotropic
porous media, where $K$, $S$ and $M$ are scalars.
Here, we explore the phenomena
occurring in the present 2D geometry with non-equivalent
properties along the $x$ and $y$ axes, 
allowing for the existence of off-diagonal effects.
\begin{figure}
\centerline{\epsfig{figure=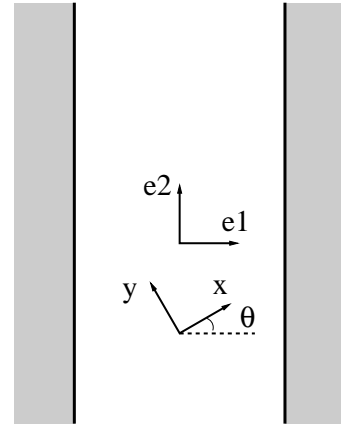}}
\vspace{.5cm}
\caption{
Top view of the slab geometry of Figure 1
}\label{f_2}\end{figure}
If the pattern is disposed at an angle $\theta$ to the 
measurement geometry, the above linear relations are best expressed
in the new system of coordinates related to the basis $(e_1,e_2)$ 
(Figs. 1 and 2).
This leads to 4x4 matrices describing the generalized conductance
of the system:
\begin{equation}
\left[
\begin{array}{l}
{\bf J}
\\ 
{\bf J_{el}} 
\end{array} 
\right]
= 
\left[
\begin{array}{lr}
{\bf K}
& 
{\bf M}
\\
{\bf M}
& 
{\bf S}
  \end{array} 
\right] 
.
\left[ 
\begin{array}{r}
-{\bf \nabla}p 
\\
-{\bf \nabla}\phi
\end{array} 
\right]   
\end{equation}
and its generalized resistance:
\begin{equation}
\left[
\begin{array}{l}
-{\bf \nabla}p 
\\ 
-{\bf \nabla}\phi
\end{array} 
\right]
= 
\left[
\begin{array}{lr}
{\bf k}
& 
{\bf m}
\\
{\bf m}
& 
{\bf s}
  \end{array} 
\right] 
.
\left[ 
\begin{array}{r}
{\bf J}
\\
{\bf J_{el}} 
\end{array} 
\right]   
\end{equation}
where the 2x2 matrices ${\bf K}$, ${\bf M}$,
${\bf S}$,
${\bf k}$,
${\bf m}$ and
${\bf s}$
are now non-diagonal
and given by 
${\bf F}={\bf R_{\theta}.{\hat F}.R_{\theta}^{-1}}$,
where ${\bf R_{\theta}}$ is the rotation matrix
between $(x,y)$ and $(e_1,e_2)$. 
\begin{equation}
{\bf R_{\theta}}=\left( \begin{array}{rr}
\cos\theta & -\sin\theta \\
\sin\theta & \cos\theta  \end{array} \right)  
\end{equation}
so that the generic matrix 
\begin{equation}
{\bf F}=\left( \begin{array}{rr}
F_{11}& F_{12}\\
F_{21}& F_{22}  
\end{array} \right)  
\end{equation}
is symmetric and given by:
\begin{equation}
{\bf F}=\left( \begin{array}{rr}
F_x\cos^2\theta +F_y\sin^2\theta & (F_x-F_y)\sin\theta \cos\theta \\
(F_x-F_y)\sin\theta \cos\theta & F_y\cos^2\theta +F_x\sin^2\theta  
\end{array} \right)  \label{def}
\end{equation}
\section{Non-diagonal effects}
Non diagonal or transverse
effects (a cause along $e_1$ induces an effect
along $e_2$ or the reverse)
occur if some of the local transverse coefficients
$K_{12}$, $M_{12}$, or $S_{12}$ are non-zero.
From the above geometric formula,
this requires $K_x \ne K_y$, $M_x\ne M_y$, or $S_x \ne S_y$,
i.e. that the anisotropy in the pattern of the plates has 
translated into different susceptibilities along the
two principal axes $x$ and $y$.

We postpone to section IV a description of a way to achieve this
asymmetry, and start with a generic description
of the effects expected 
if local anisotropy is present.

\subsection{Open homogeneous geometries}

In an open homogeneous geometry (no walls in Figure 1),
the two simplest situations are the following:

{\em No pressure differences - }
Then an electric field ${\bf E}$ generates an electric current 
${\bf J_{el}}={\bf S.E}$ and an {\em electro-osmotic} flow
${\bf J}={\bf M.E}$, at finite angles with respect to the applied
field ${\bf E}$, if ${\bf S}$ and ${\bf M}$ are non-diagonal.
Explicit ways to obtain non-diagonal electro-osmotic flows 
were proposed in
\cite{Ajd1}, and are revisited in section IV within a simple 
lubrication picture.

{\em No electrostatic potential differences - }
If all potential differences are short-circuited by an appropriate
design of electrodes connected by low-resistance wiring,
then a pressure gradient creates not only 
a hydrodynamic flow ${\bf J}={\bf K.}(-\nabla p)$, but also
an electric current ${\bf J_{el}}={\bf M.}(-\nabla p)$ through
the wiring circuit. 
This current in a simple 1D geometry is called 
the streaming current. Both hydrodynamic and electric currents
are again generically not aligned with the applied pressure gradient.

As this remains formal, 
geometries are now considered where the fluid is confined 
to a rectangular channel.

\subsection{Channel with passive walls}
Take a channel of length $L$ along $e_2$
and width $d$ along $e_1$, bounded by impermeable walls.
We suppose here that no electric
current can flow from one side to the other,
so both
fluxes in direction $e_1$ are on average zero:
$J_1=0$ and $J_{{\rm el}1}=0$.
Potential differences between the two walls can however be measured
using a set of electrodes connected by a high resistance in series with 
a voltmeter (Fig. 3).
In this geometry 
it is straightforward to
quantify the effects generated by a forcing along the length
of the channel (direction $e_2$).

\subsubsection{Pressure-driven effects } 
Suppose a pressure drop
$\Delta p_2$ is applied along the channel so that the average value of 
$\partial_2 p$ is $\Delta p_2/L$. We furthermore
consider that there is no external electrical path 
connecting the two ends of the channel 
(in particular the fluid flow does not carry in or out
a net current - no streaming current  carried by the convection
of the neutral fluid at the inlet and outlet of the channel-) 
so that $J_{{\rm el} 2}=0$:
any streaming current (along the walls) is compensated by a back-current
in the bulk.

The application of the pressure drop $\Delta p_2$ then 
results in longitudinal effects along $e_2$:
a fluid flow $J_2$ and a potential drop
$\Delta \phi_2$ given by:
\begin{eqnarray}
(-\Delta p_2/L) &=& k_{22} J_2 \\
(-\Delta \phi_2/L)&=& m_{22} J_2
\end{eqnarray}
The potential drop $\Delta \phi_2=(m_{22}/k_{22})\Delta p_2$
is the classical ``streaming potential'', proportional
in the simplest cases to the average surface potential of the 
walls. The flow $J_2$ is given by a Darcy law modified
by electric effects: $J_2= k_{22}^{-1}(-\Delta p _2/L)$.

More interestingly, transverse effects are also generated
in the $e_1$ direction: a pressure difference 
$\Delta p_1$ and a potential drop $\Delta \phi_1$:
\begin{eqnarray}
(-\Delta p_1/d) &=& k_{12} J_2 = (k_{12}/k_{22}) (-\Delta p _2/L) \\
(-\Delta \phi_1/d)&=& m_{12} J_2 = (m_{12}/k_{22}) (-\Delta p _2/L)
\end{eqnarray}
The pressure difference $\Delta p_1$ exists if $k_{12} \neq 0$,
in which case it indicates that recirculation is taking place in the
$e_1$ direction: the pressure drop along $e_2$ entrains fluid in the $e_1$ 
direction, which, due to the presence of the walls, then has to recirculate.
A certain amount of shearing and mixing is thus induced.

The electric potential difference $\Delta \phi_1$ is a 
``transverse streaming potential `` proportional to the
non-diagonal electrokinetic coefficient $m_{12}$.
It is 
a transverse electric measure of the
hydrodynamic flow along $e_2$. Note that, 
even if the local off-diagonal coupling $m_{12}/k_{22}$
is large,
this potential drop is intrinsically
smaller than the streaming potential along the channel by a factor
$d/L$ due to the geometry of the system.
\begin{figure}
\centerline{\epsfig{figure=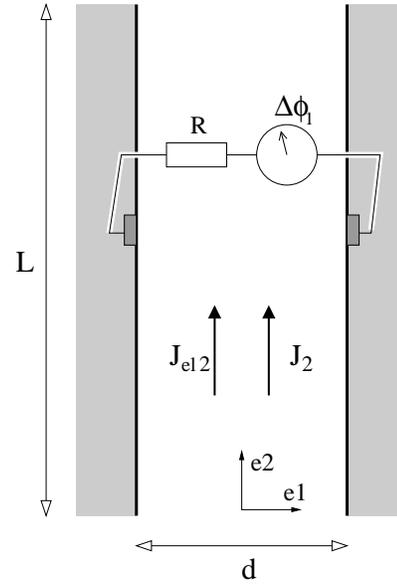}}
\vspace{.2cm}
\caption{
Passive walls: Neither hydrodynamic flow nor electric current
is possible along the $e_1$ axis. A potential difference
$\Delta \phi_1$ can be measured using a high input impedence voltmeter.
}
\end{figure}
\subsubsection{Electrically-driven effects } 
Suppose now that an electric current $J_{{\rm el} 2}$,
or an average electric field $E_2$, is applied along a
channel connected to large open reservoirs
so that $\Delta p_2 =0$. 

This yields classical effects along the channel, which are 
simply Ohm's law and the ``electro-osmotic flow'':
a net flow $J_2$ induced by the electric field:
\begin{eqnarray}
(-\Delta p_2/L) &=& k_{22} J_2 + m_{22} J_{{\rm el}2} = 0 \\
(-\Delta \phi_2/L)&=& m_{22} J_2 + s_{22}J_{{\rm el}2}= E_2 
\end{eqnarray}
The electro-osmotic flow is thus
$J_2=- (m_{22}/k_{22})J_{{\rm el}2}= 
-\frac{m_{22}}{k_{22}s_{22}-m_{22}^2} E_2$
while the resistivity of the channel is given by
$E_2= (s_{22}-m_{22}^2/k_{22}) J_{{\rm el}2}$. 

In addition, transverse effects are generated
in the perpendicular $e_1$ direction: a pressure difference 
$\Delta p_1$ and a potential drop $\Delta \phi_1$.
\begin{eqnarray}
(-\Delta p_1/d) &=& k_{12} J_2 + m_{12}J_{{\rm el}2} 
=\frac{m_{12}k_{22}-k_{12}m_{22}}{s_{22}k_{22}-m_{22}^2} E_2  \\
(-\Delta \phi_1/d)&=& m_{12} J_2 +s_{12}J_{{\rm el}2} = 
\frac{s_{12}k_{22}-m_{12}m_{22}}{s_{22}k_{22}-m_{22}^2} E_2 
\end{eqnarray}
Equation (19) indicates that the electric field $E_2$ induces
an electro-osmotic flow along $e_1$ (if $m_{12}\neq 0$),
which due to the presence of the walls, leads to a pressure difference 
$\Delta p_1$ that induces recirculation of the fluid along $e_1$. 
Again this induces shear and favors mixing.
This effect is due both  to the non-diagonal permeation ($k_{12}$)
and to transverse electro-osmosis per se ($m_{12}$).

Equation (20) describes an ``electrokinetic Hall effect'':
an electric field applied along $e_2$ results in an electrostatic
potential difference along $e_1$: it consists of a term due
to the anisotropic conductivity ($s_{12}$) and to transverse streaming
current ($m_{12}$).
\subsection{Comments}

Many other geometries can actually be envisaged. 
In Appendix 1 we consider situations where
electrodes are embedded in the walls, which permits, for example, the 
 generation of flow along the channel from transverse potential differences
between the walls. Let us stress here a few general points:\\
- Evaluation of the effect is easy only if the geometry
(imposed boundary conditions) permits homogeneous solutions
for the gradients and currents. Otherwise one has to solve
the 2D conservation equations for the currents.\\
- This obviously would also be the case
with heterogeneous patterns on the plates.
We will return to this in the discussion section.\\
- A more serious (less obvious) problem has to do with
electrodes: the specificities of electrochemistry at their surface
may require a description in terms of the fluxes of the various
ions rather than the 
simple description in terms of an electric current of Appendix 1,
in addition to the many experimental problems involved 
(surface corrosion, generation of gas bubbles, etc ...).\\
- In the channel geometries considered here, modifying or controlling
the electrical 
connection between the walls affects the longitudinal coefficients $K_{22}, M_{22}, S_{22}$
(if off-diagonal coefficients are non-zero). 

Clearly several coefficients describe the various couplings allowed by
symmetry in this linear theory. 
To give quantitative estimates for these coefficients and thus for the 
corresponding effects one has to deal with notoriously difficult
(even at low Reynolds number) electro-hydrodynamic calculations.
However a useful guide can be developed using the lubrication approximation,
as shown in the next section.
\section{Specific Calculations within lubrication approximation}

\subsection{Geometry and approximations}
In this section, modulating the shape and the charge density
on the plates is proposed as an efficient patterning 
to obtain transverse effects.
Explicit calculations of the matrix
coefficients introduced in section II are performed,
which provide an estimate of the off-diagonal effects as listed in section III.
For the analysis to remain simple, I 
focus on the specific case where 
the charge and/or surface pattern is periodic along the direction $x$
and invariant along the perpendicular direction $y$.
To set notations, the channel has a local thickness
$h(x)= h_0 +\delta h(x)$, 
and modulated charge densities $\sigma_1(x)$ on the bottom plate
and $\sigma_2(x)$ on the top one (Figure 4).
Further, the following assumptions are made:

(i) the modulation wavelength is larger than the gap so that
lubrication approximation holds \cite{Hap}.

(ii) the ionic strength is high enough for the typical
gap thickness $h_0$ to be much larger than the 
Debye length $\lambda_D=\kappa^{-1}$,

(iii) the surface potentials (or charge densities) are weak enough
for the double layers to be well described
by the Debye-H\"uckel theory. All effects are calculated to first order in
these surface charge densities, so that terms proportional
to products of $\sigma_i$ are neglected altogether.
This restriction comes in addition to (and is distinct from) the fact
that we focus on the linear response of fluxes to applied fields.

Although these assumptions limit the quantitative precision of the
results, they permit analytic calculations
and thus a clear discussion of important qualitative features
at a reasonably simple level.

\begin{figure}
\centerline{\epsfig{figure=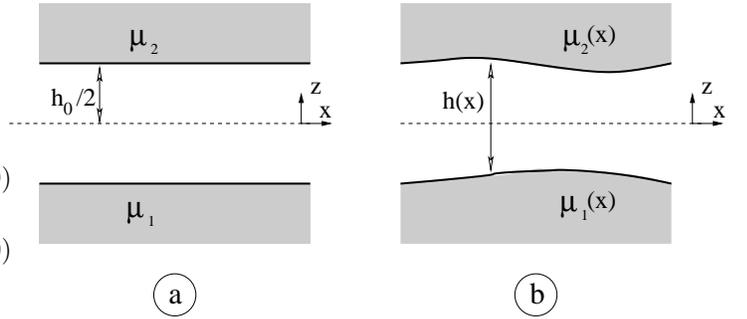}}
\vspace{.5cm}
\caption{
Geometry considered in Section IV : a) case of a uniform flat
system, b) the slip parameters describing
electrokinetic effects and the thickness vary along $x$ only.
}\label{f_4}\end{figure}

\subsection{Flat and uniform surfaces}

To get a physical insight,
it is useful to start with the simple geometry of uniform flat surfaces (Fig. 4a):
$h(x)=h_0$, $\sigma_1$ and $\sigma_2$ constant.
The computation of electrokinetic effects is then 
simple (given (ii) and (iii) \cite{And1}), and 
can be found in textbooks (e.g. \cite{Hun1}).\\

{\em Electro-osmosis - }
If an electric field $E$ is present in the channel,
it will exert a drag on the thin charged Debye layer in the vicinity 
of the surfaces. The actual no-slip boundary condition
for the solvent flow on the real surface of the plates
can then be replaced by an electro-osmotic slip 
velocity $v_i=-\mu_i E$ on plate $i$, where   
$\mu_i= \sigma_i\lambda_D/\eta$, and $\eta$ is the viscosity of water.
This leads to a simple flow profile
$v_{EO}(z)=-\mu_1 E (1-2z/h_0) -\mu_2 E (1+2z/h_0)$ and
a net electro-osmotic flow through the channel
$J_{EO}=-(\mu_1+\mu_2)h_0E/2$.\\

{\em Poiseuille flow - }
In addition there is naturally the pressure driven flow 
$v_{Poiseuille}=(z^2-h_0^2/4)\partial_xp/2\eta$
which gives a current $J_{Poiseuille}=-h_0^3\partial_xp/12\eta$.\\

{\em Flow-induced current - }
This pressure-driven flow induces transport of the charged fluid of the 
Debye layers. If the local shear rate in the vicinity
of the $i$-th plate is ${\dot \gamma}$,
the induced current is (per unit length in the $y$ direction)
$J_{{\rm el} i}= -\sigma_i \lambda_D {\dot \gamma}=-\eta \mu_i {\dot \gamma}$.
Given that ${\dot {\gamma}}=- h_0 \partial_xp/2\eta$,
this leads to a total
current $J_{{\rm el} Flow-induced}=(\mu_1+\mu_2)h_0\partial_xp/2$.
Note that in principle, electro-osmotic flows also induce convective 
transport of the charged regions, but this effect
is then clearly proportional to $\sigma_i\sigma_j$ and neglected
here along (iii).\\

{\em Ohm's law - }
Of course the flow-induced current is often a minor correction to the
main conduction current described by Ohm's law 
$J_{{\rm el} Ohm}=\sigma_{el}h_0E$.\\

Gathering all contributions, we are thus led to a scalar
(in this isotropic case) version of equations (1) and (2):
\begin{eqnarray}
J &=& \frac{h_0^3}{12\eta} .(-{\bf \nabla}p) - \,  h_0\frac{\mu_1+\mu_2}{2}
. (-{\bf \nabla}\phi)\\
J_{el} &=& -h_0\frac{\mu_1+\mu_2}{2} . (-{\bf \nabla}p) + 
\sigma_{el}h_0 .(-{\bf \nabla}\phi).
\end{eqnarray}
The symmetry of the Onsager matrix is here clear.
\subsection{Fields and currents along the modulation direction $x$}

Returning to a channel periodically modulated along $x$,
I consider first the case where the fluxes and gradients are applied
along this axis. In this geometry, the currents $J_x$ and $J_{{\rm el} x}$
are constants,
whereas the local values of the pressure gradient $\partial_xp(x)$ 
and electric field $E_x(x)$ vary
(although slowly enough in the lubrication approximation
to consider them independent of $z$).

In the lubrication picture we can transcribe the results 
of the previous subsection:
\begin{eqnarray}
J_x &=& \frac{h^3(x)}{12\eta} .(-\partial_x p(x))  
- \, h(x)\frac{\mu_1+\mu_2}{2}(x)
. E_x(x)\\
J_{{\rm el} x} &=& -h(x)\frac{\mu_1+\mu_2}{2}(x) . (-\partial_x p(x)) + 
\sigma_{el}h(x) .E_x(x)
\end{eqnarray}
Then, neglecting terms proportional to the product of $\sigma_i$s
(or $\mu_i$s), and performing integrals along $x$ over 
a period of the modulation, gives

\begin{equation}
\left(\! 
\begin{array}{r}
\left<-\partial_x p\right>\\\left<-\partial_x \phi \right> \end{array} 
\!\right) 
=\frac{12 \eta}{\sigma_{el}}
\left( 
\begin{array}{lr}
\sigma_{el}\left<\frac{1}{h^3}\right>& 
\left<\frac{\mu_1+\mu_2}{2h^3}\right> \\
\left<\frac{\mu_1+\mu_2}{2h^3}\right>& \frac{1}{12\eta}
\left<\frac{1}{h}\right>  \end{array} 
\right) 
\left(\! 
\begin{array}{l}
J_x \\ J_{{\rm el} x}\end{array} 
\!\right)  
\end{equation}
where $\left<f\right>$ is the average of the function $f(x)$ 
over a period.
With the same approximations, inversion yields:

\begin{equation}
\left( \!
\begin{array}{l}
J_x \\ J_{{\rm el} x}\end{array} 
\!
\right)
= 
\Delta_x
\left( 
\begin{array}{lr}
\frac{1}{12\eta}\left<\frac{1}{h}\right>
& 
-\left<\frac{\mu_1+\mu_2}{2h^3}\right> 
\\
-\left<\frac{\mu_1+\mu_2}{2h^3}\right>
& 
\sigma_{el} \left<\frac{1}{h^3}\right>
  \end{array} 
\right) 
\left( 
\!
\begin{array}{r}
\left<-\partial_x p\right>\\\left<-\partial_x \phi \right> 
\end{array} 
\!
\right)  
\end{equation}
with $\Delta_x^{-1}=
\left<\frac{1}{h^3}\right>
\left<\frac{1}{h}\right>$. 

\subsection{Fields and currents along the perpendicular direction $y$}

I now take fields and flows along the perpendicular direction $y$.
In an infinite geometry, the electric field $E_y$ and the pressure gradient
$\partial_yp$
are independent of $x$, whereas the currents are modulated
along this direction.
\begin{eqnarray}
J_y(x) &=& \frac{h^3(x)}{12\eta} .(-\partial_y p) 
- \, h(x)\frac{\mu_1+\mu_2}{2}(x)
. E_y\\
J_{{\rm el} y}(x) &=& -h(x)\frac{\mu_1+\mu_2}{2}(x) . (-\partial_yp(x)) + 
\sigma_{el}h(x) .E_y
\end{eqnarray}
Averaging over a period along $x$ here gives:

\begin{equation}
\left( \!
\begin{array}{l}
\left<J_y\right> \\ \left<J_{{\rm el} y}\right>\end{array} 
\!\right)
= 
\left( 
\begin{array}{lr}
\frac{1}{12\eta}\left<h^3\right>
& 
-\left<\frac{\mu_1+\mu_2}{2}h\right> 
\\
-\left<\frac{\mu_1+\mu_2}{2}h\right> 
& 
\sigma_{el}\left<h\right>
  \end{array} 
\right) 
\left(\! 
\begin{array}{r}
-\partial_y p\\-\partial_y \phi  \end{array} 
\!\right)   
\end{equation}
or upon inversion
\begin{equation}
\left(\! 
\begin{array}{r}
-\partial_y p\\-\partial_y \phi  \end{array} 
\!\right) 
=\Delta^{-1}_y
\frac{12\eta}{\sigma_{el}}
\left( 
\begin{array}{lr}
\sigma_{el}\left<h\right>
& 
\left<\frac{\mu_1+\mu_2}{2}h\right> 
\\
\left<\frac{\mu_1+\mu_2}{2}h\right> 
& 
\frac{1}{12\eta}\left<h^3\right>
\end{array} 
\right) 
\left(\! 
\begin{array}{l}
\left<J_y\right> \\ \left<J_{{\rm el} y}\right>\end{array} 
\!\right)  
\end{equation}
with $\Delta_y^{-1}=\left<h^3\right>^{-1}
\left<h\right>^{-1}$.

\subsection{Summary for arbitrary modulations}

We can now write down
the local coefficients of the response matrix 
to make the anisotropy in $(x,y)$ explicit.
The coefficient of the permeability matrices (section II) 
giving fluxes as functions of gradients are:
\begin{eqnarray}
K_x = \frac{1}{12\eta} \left<1/h^3\right>^{-1} \,;&\,\, 
K_y = \frac{1}{12\eta} \left<h^3\right> \\
S_x = \sigma_{el} \left<1/h\right>^{-1} \,;& \,\,
S_y = \sigma_{el} \left<h\right> \\
M_x = -
\frac{
\left<\frac{\mu_1+\mu_2}{2h^3}\right>
}{
\left<\frac{1}{h^3}\right>
\left<\frac{1}{h^1}\right>
} 
\,;&\,\,
M_y = -\left<\frac{\mu_1+\mu_2}{2}h\right>
\end{eqnarray}
The anisotropy appears in a similar form in the 
``resistance matrix'' :
\begin{eqnarray}
k_x = 12\eta \left<1/h^3\right> \,;&\,\, 
k_y = 12\eta \left<h^3\right>^{-1} \\
s_x = \frac{1}{\sigma_{el}} \left<1/h\right> \,;& \,\,
s_y = \frac{1}{\sigma_{el}} \left<h\right>^{-1} \\
m_x = \frac{6\eta}{\sigma_{el}}
\left<\frac{\mu_1+\mu_2}{h^3}\right>
\,;&\,\,
m_y =
\frac{6\eta}{\sigma_{el}}
\frac{
\left<(\mu_1+\mu_2)h\right>
}{
\left<h\right>\left<h^3\right>
}
\end{eqnarray}
The coefficients describing off-diagonal effects (section III)
are then obtained from these matrices expressed
in the  $(e_1,e_2)$ frame 
(using equations (11) and (12)) .

\subsection{Discussion}

\begin{figure}[t]
\vspace{-.3cm}
\centerline{\epsfig{figure=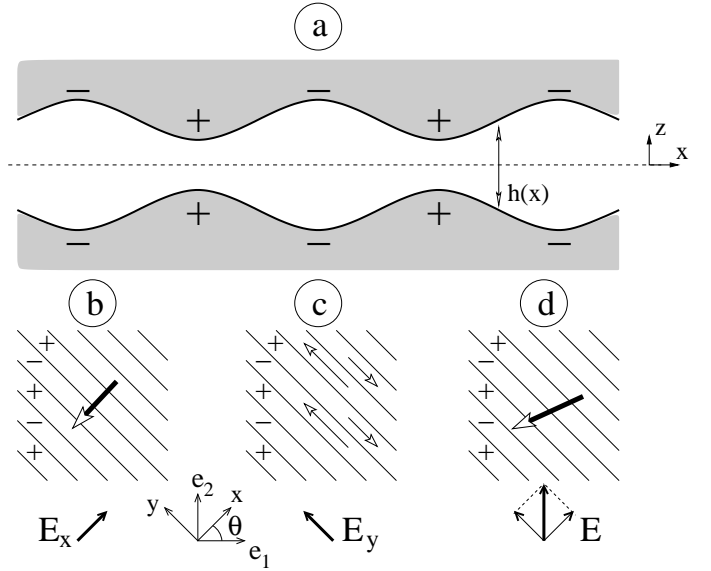}}
\vspace{.5cm}
\caption{
Generation of off-diagonal effects (here electro-osmosis) from a sinusoidal modulation
of shape and charge densities. The case depicted here (a) corresponds to 
$\alpha >0$, $\delta\mu >0$, $\Theta=\pi/2$, and a zero average charge density $\mu_0=0$.
A schematic top view is then used in (b),(c), and (d), with $\theta=\pi/4$.
(b): The field $E_x$
generates a flow along $x$ resulting from the opposite action of the narrow
positively charged sections 
(tending to slip the fluid backwards along $x$) and of the wider negatively charged
sections (which induce a slip
in the positive direction). The thin sections 
dominate because of their stronger hydrodynamic resistance
so the net flow is opposite to the applied field.
(c): A field $E_y$ applied along $y$ induces a stratified flow with thin and wide sections again
pushing in opposite directions. As the field is uniform, the slip velocities
in the two directions are of similar amplitudes.
Thus the net flow is dominated by the wider sections, and points
in the direction of the applied field. (d): Thanks to the linearity of the problem,
the net flow created by a field $E$ applied along $e_2$ is obtained
by superimposing 
those obtained in (b) and (c), resulting here in a dominant transverse component.
}\label{f_sch}
\end{figure}

To investigate semi-quantitatively the consequences
of the formulas derived above,
a simple but 
insightful
example is sinusoidal modulations
of the shape and charge density (Figure 5a)
\begin{eqnarray}
h(x)=h_0 (1 +\alpha \cos(qx))\\
\mu_1+\mu_2 = 2  (\mu_0 +\delta\mu \cos(qx+\Theta)).
\end{eqnarray}
Analytical results take a simple form (see Appendix 2)
in the limit of weak 
modulation amplitudes for the shape, which allows
to calculate the result
through an expansion in $\alpha$. From this and from inspection
of the formulae of the subsection above,
a few important points can be derived:
\begin{itemize}

\item a simple modulation of the charge pattern is not sufficient to
induce off-diagonal effects.
This is the consequence of the linear description in terms of 
the surface charge densities and of the $+/-$ symmetry (see e.g. \cite{Ajd1}
for a discussion of this)

\item a modulation of the shape at homogeneous charge density
is enough to produce off-diagonal effects
for all phenomenologies ( permeation $k_{12}\neq0$, electro-osmosis
and streaming potential/current $m_{12} \neq 0$, conductance $s_{12}\neq 0$).
However for a slight modulation $\delta h(x)$ around a mean $h_0$,
the coupling coefficients will be proportional to $(\delta h /h_0)^2$.

\item a correlated coupling of the charge pattern and of the shape
leads to stronger $O(\delta h/ h_0)$ amplitudes for the 
off-diagonal electro-hydrodynamic couplings $m_{12}$ and $M_{12}$.
These off-diagonal couplings are proportional to $\delta\mu$, and 
dominate the longitudinal ones if $\delta\mu/\mu_0 \gg 1$.
This synergy between shape and charge modulation
stems from the shape-induced symmetry breaking between plus and minus charges,
and is described at length in \cite{Ajd1} for the specific case of electro-osmosis.
A shematic description is given in figure \ref{f_sch}.

\item extrapolation of these results to the case $\alpha \rightarrow 1$
semi-qualitatively suggests that the off-diagonal components can be
of the same order than the diagonal one. Favorable
ingredients are a marked shape modulation
($\delta h /h \sim 1$), and a rather contrasted charge modulation ($\delta\mu/\mu_0 \sim 1$).

\end{itemize}

Indeed, the results of Appendix 2 show that
the off-diagonal
matrix coefficients ($k_{12},m_{12},s_{12},K_{12},M_{12},S_{12}$)
are smaller by a factor $\sim \alpha^2$
than their diagonal equivalents if only the shape is modulated
($\delta\mu=0$). If in addition the charge density is commensurately
modulated the off-diagonal electro-hydrodynamic
coefficients $m_{12}$ and $M_{12}$ are increased by a factor
$\alpha\frac{\delta\mu}{\mu_0}\cos(\Theta)$ relative to their diagonal equivalents.

\section{Conclusion}

\begin{figure}
\centerline{\epsfig{figure=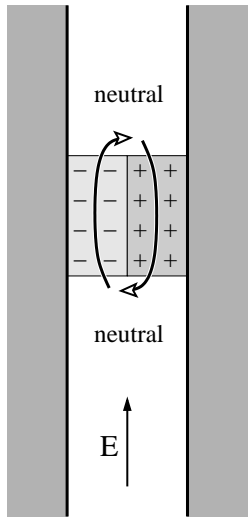}}
\vspace{.5cm}
\caption{
A simple heterogeneous geometry to create a vortex (top view).
No shape modulation, and a simple charge pattern : 
a positively charged zone aside a 
a negatively charged one (ideally both plates are patterned in the same
way, i.e. floor and ceiling).
The rest of the capillary is neutral. The electro-osmotic slip 
generated by an electric field along the channel
leads to a recirculation vortex as shown.
}\label{f_vor}
\end{figure}

We have shown that various off-diagonal effects can be generated
using micro-patterned geometries allowing for various functionalities:
transverse pumping, mixing, flow detection, etc... The
corresponding  couplings can locally be of same order
as the usual (diagonal) electro-kinetic coefficients (sometimes larger). Their actual 
global value
depends on the geometry of the given device (through geometrical ratios 
such as e.g. $L/d$).  A detailed analysis of the flow geometries
(variations in the third ($z$) direction)
and thus of the mixing capabilities will be published elsewhere.

Realization of charge-patterned surfaces at the 10-micron scale
has already been experimentally explored in simple geometries \cite{Str1}.
Its combination with shape modulations is clearly within 
the range of current microfabrication technologies, and is the topic
of ongoing studies.

Naturally, more efficient devices for pumping, mixing or flow
detection should be realizable by playing with a larger class 
of patterns, including in particular inhomogeneous ones.
A simple example is schematized in Figure \ref{f_vor} where
a vortex can be created by application of an electric field.
Additional features or control can result
from the  increase of the third dimension : dealing with thicker
fluid layers allows, for example, to produce rolls with axes parallel to the surface
\cite{Ajd1,Str1}. 
Eventually let us stress that although the focus of the present paper is on steady-state effects
obtained with d.c. fields, interesting effects are also expected using a.c. fields
and arrays of microelectrodes \cite{Fuh1,Mul1,Ajd2,Bro1}.
This great diversity of geometries imposes a hand-in-hand
development of theoretical proposals and experimental realizations that we wish to
pursue in the coming years.\\

{\em Acknowledgments: This work owes much to a continuous interaction
with Abe Stroock, and I thank him for his permanent support and for many useful suggestions. 
Related discussions with Y. Chen, D. Long, A. P\'epin and
 P. Tabeling are gratefully acknowledged.}


\newpage
 
\section*{Appendix 1: Walls shortcut by connected electrodes}

We now consider the geometry of section III, 
but with the walls
perpendicular to $e_1$ now covered by 
connected electrodes, so as to shortcut any electrostatic
potential difference between the two walls.

\begin{figure}
\centerline{\epsfig{figure=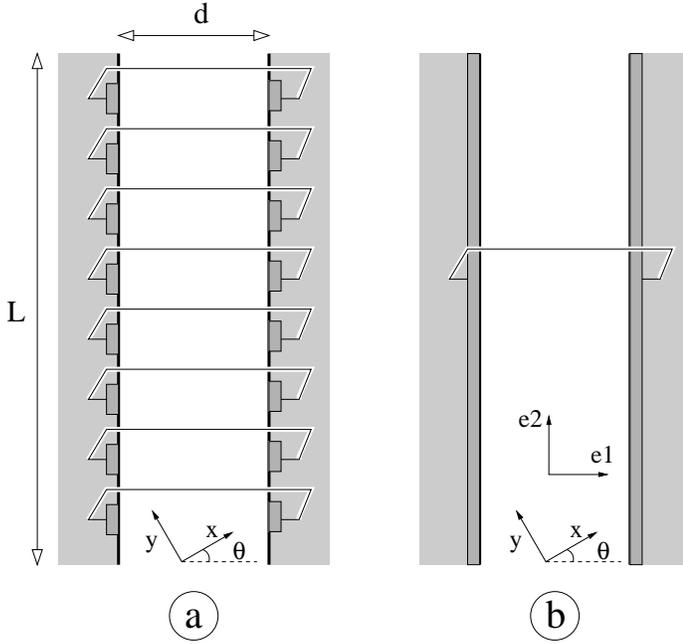}}
\vspace{.5cm}
\caption{
(a): connected electrodes; (b):
 the continuous electrodes short-circuit potential
differences along the channel}
\label{f_bi}
\end{figure}

Our picture here is 
very simplistic as we do not want to enter the detail
of electrochemical reactions. Let us only make the following 
distinction (Figure \ref{f_bi}). First, case (a), we will consider
that these walls actually consist of
a series of electrode pairs not connected between them,
so that although the potential drop
is on average zero in the $e_1$ direction,
potential differences can nevertheless exist along the $e_2$ direction.
In the second simplistic case (b), the walls are coated with 
continuous electrodes in which currents can circulate
also along the $e_2$ direction so that the electrostatic potential
is essentially constant $\Delta \phi_1=\Delta \phi_2=0$.

\subsection{Series of connected electrodes}

{\em Pressure driven flow - }
In this geometry (Figure 7a) ($J_{{\rm el}2}=0$, $J_1=0$, $\Delta \phi_1=0$), 
an applied pressure difference along the channel
$\Delta p_2 \neq 0$ results in a streaming
potential $\Delta \phi_2$ that differs from the
one obtained with passive walls:
\begin{equation}
\Delta \phi_2= \frac{s_{11}m_{22}-s_{21}m_{12}}{s_{11}k_{22}-m_{12}^2}
\Delta p_2
\end{equation}

In addition, this pressure drop creates transverse
effects: a pressure drop 
$\Delta p_1$ along the $e_1$ axis that induces recirculation,
as well as a {\em transverse streaming current} $I_1=L J_{{\rm el}1}$
given by:
\begin{eqnarray}
-\Delta p_1&=&  \frac{d}{L} \frac{s_{11}k_{12}-m_{11}m_{12}}
{s_{11}k_{22}-m_{12}^2 } \Delta p_2
\\
I_1&=& L J_{{\rm el}1} = \frac{m_{12}}
{s_{11}k_{22}-m_{12}^2 } \Delta p_2
\end{eqnarray}

{\em Electrically driven flow - }
In the case where no pressure drop exists between the two
ends of the channel
($\Delta p_2 = 0$), effects can be generated
by applying an electric current
$J_{{\rm el}2} \neq 0$ (or an electric field $E_2$ along $e_2$
(due to the short-circuited walls we have $J_1=0$, $\Delta \phi_1=0$).
The applied field creates an electro-osmotic flow 
along the channel:
\begin{equation}
J_2=\frac{M_{22}K_{11}-K_{21}M_{12}}{K_{11}}E_2
\end{equation}
that is different from the case where the walls were not connected
(equation below (18)), if off-diagonal coefficients are non-zero.
The electric conduction law is here 
$J_{{\rm el}2}=\frac{S_{22}K_{11}-M_{12}^2}{K_{11}}E_2$.
{\em Transverse electro-osmosis} results in a pressure difference
$\Delta p_1 = d (M_{12}/K_{11})E_2$ and a total current intensity
$I_1=LJ_{{\rm el}1}= L\frac{S_{12}K_{11}-M_{12}M_{11}}{K_{11}}E_2$
through the connecting wires.

\subsection{Continuous conducting electrodes}

Due to the presence of the electrodes,
any potential difference between the entrance and the exit of the channel
is short-circuited (i.e. a backward current runs along the electrodes
in the $e_2$ direction) so that $\Delta \phi_2=0$ (Figure 7b).

{\em Pressure driven flow - }
A pressure drop along the channel ($\Delta p_2\neq 0$)
in this situation 
($\Delta\phi_2=0$, $J_1=0$, $\Delta \phi_1=0$) again creates
 a transverse 
pressure drop across the channel 
$\Delta p_1$ synonymous of recirculation along the $e_1$ axis,
as well as a {\em transverse streaming current} $I_1=L J_{{\rm el}1}$
given by:
\begin{eqnarray}
-\Delta p_1&=& - \frac{d}{L} \frac{K_{12}}{K_{11}} \Delta p_2
\\
I_1&=& - \frac{M_{12}K_{11}-M_{11}K_{12}}{K_{11}}\Delta p_2
\end{eqnarray}

{\em Transverse electro-osmotic pumping - }
But we can also consider the situation where an electric current
$J_{{\rm el}1}$ is applied between the two walls.
Formally, this case ($\Delta \phi_2=0$, $J_1=0$) allows
electro-osmotic pumping in the $e_2$ direction, even in the absence
of a driving pressure gradient in that direction ($\Delta p_2=0$).
The resulting flow is 
\begin{equation}
J_2= (\frac{M_{21}K_{11}-K_{21}M_{11}}{K_{11}})E_1
\end{equation}
or $J_2
=(\frac{M_{21}K_{11}-K_{21}M_{11}}{K_{11}S_{11}-M_{11}^2}) J_{{\rm el}1}
$.
This is simply a manifestation of the transverse electro-osmosis
mentioned in the first subsection
of this section. It allows the design of lateral electro-osmotic
pumps that do not require a global potential difference between
the entrance and the exit of the channel (i.e. $\Delta \phi_2$ proportional
to $L$) but rather local potential drops ($\Delta \phi_1$ proportional
to $d$).

\section*{Appendix 2: Sinusoidal
modulations of shape and charge densities}

We propose here explicit formulae for the various components
of the conductance and resistance matrices, obtained from the 
lubrication approximation in section (IV), and summarized
in IV-E, for the particular case of
sinusoidal modulations of the height of the channel
$h=h_0(1+\alpha \cos(qx))$ and of the charge densities
$(\mu_1+\mu_2)/2=\mu_0 + \delta\mu \cos(qx+\Theta)$.

The lubrication approximation requires that $\alpha h_0 q \ll 1$.
We here present explicit formulae that correspond to an expansion 
in $\alpha$ up to order $\alpha^2$, assuming explicitely
$\alpha \ll 1$. This provides a useful guide.  
The results are for the conductance matrix:

\begin{eqnarray}
K_x = \frac{h_0^3}{12\eta} (1-3\alpha^2) \,;&\,\, 
K_y = \frac{h_0^3}{12\eta} (1+\frac{3}{2}\alpha^2) \\
S_x = \sigma_{el}h_0 (1-\frac{1}{2}\alpha^2) \,;& \,\,
S_y = \sigma_{el}h_0 \\
M_x = -\mu_0h_0 &(1-\frac{3}{2}\alpha\frac{\delta\mu}{\mu_0}
\cos(\Theta)-\frac{1}{2}\alpha^2)
\\
M_y = -\mu_0h_0& \!\!\!\!\!\!\!\!\!(1+\frac{1}{2}\alpha\frac{\delta\mu}{\mu_0}\cos(\Theta))\,\,\,\,\,
\end{eqnarray}

and for the resistance matrix

\begin{eqnarray}
k_x = \frac{12\eta}{h_0^3} (1+3\alpha^2)\,;&\,\, 
k_y =  \frac{12\eta}{h_0^3}(1-\frac{3}{2}\alpha^2) \\
s_x = \frac{1}{\sigma_{el}h_0}(1+\frac{1}{2}\alpha^2)  \,;& \,\,
s_y = \frac{1}{\sigma_{el}h_0} \\
m_x = \frac{12\eta\mu_0}{\sigma_{el}h_0^3}&
(1-\frac{3}{2}\alpha\frac{\delta\mu}{\mu_0}\cos(\Theta)+3\alpha^2)\\
m_y = \frac{12\eta\mu_0}{\sigma_{el}h_0^3}&
(1+\frac{1}{2}\alpha\frac{\delta\mu}{\mu_0}\cos(\Theta)-\frac{3}{2}\alpha^2)
\end{eqnarray}

Thus, the ratios of transverse (12) to longitudinal (11 or 22)
coefficients are

\begin{eqnarray}
k_{12}/k_{11}= k_{12}/k_{22} 
\simeq & \frac{9}{2}\alpha^2\, \sin\theta\cos\theta\\
s_{12}/s_{11}= s_{12}/s_{22} 
\simeq & \frac{1}{2}\alpha^2\, \sin\theta\cos\theta\\
m_{12}/m_{11}= m_{12}/m_{22} 
\simeq & (-2\alpha\frac{\delta\mu}{\mu_0}\cos\Theta
+\frac{9}{2}\alpha^2) \, \sin\theta\cos\theta\\
K_{12}/K_{11}= K_{12}/K_{22} 
\simeq & -\frac{9}{2}\alpha^2\, \sin\theta\cos\theta\\
S_{12}/S_{11}= S_{12}/S_{22} 
\simeq & -\frac{1}{2}\alpha^2\, \sin\theta\cos\theta\\
M_{12}/M_{11}= M_{12}/M_{22} 
\simeq & (-2\alpha\frac{\delta\mu}{\mu_0}\cos\Theta
-\frac{1}{2}\alpha^2) \, \sin\theta\cos\theta
\end{eqnarray}
where terms of order $\alpha^2(\frac{\delta\mu}{\mu_0})^2$ have been omitted
in the equations for $m_{12}$ and $M_{12}$. 
Actually, if the average charge is almost zero so that $\delta\mu/\mu_0$
is very large, then in the limit $\alpha\frac{\delta\mu}{\mu_0}\gg1$,
equations (56) and (59) become:
\begin{equation}
m_{12}/m_{11}
\simeq M_{12}/M_{11} 
\simeq 
\frac{4 \sin\theta\cos\theta}{3\cos^2\theta - \sin^2\theta}
\end{equation}
so that longitudinal effects are totally dominated by transverse
ones for a ``magic angle'' $\theta=\pi/3$.


\end{document}